\documentclass[twocolumn,prl,showpacs,preprintnumbers]{revtex4}%
\usepackage{amssymb}
\usepackage[dvips]{graphicx}
\usepackage{dcolumn}
\usepackage{bm}
\usepackage{amsmath}
\usepackage{amsfonts}
\usepackage{revsymb}%
\begin{document}
\title{Interferometric determination of the $s$- and $d$-wave scattering amplitudes
in $^{87}$Rb}
\author{Ch.~Buggle}
\author{J.~L\'{e}onard}
\author{W.~von Klitzing}
\author{J. T. M.~Walraven}
\affiliation{FOM Institute for Atomic and Molecular Physics\mbox{,} Kruislaan 407\mbox{,}
1098 SJ Amsterdam\mbox{,} The Netherlands and Van~der~Waals-Zeeman Institute
of the University of Amsterdam\mbox{,} Valckenierstraat 65/67\mbox{,} 1018 XE
The Netherlands}
\date{\today }

\begin{abstract}
We demonstrate an interference method to determine the low-energy elastic
scattering amplitudes of a quantum gas. We linearly accelerate two ultracold
atomic clouds up to energies of $1.2$ mK and observe the collision halo by
direct imaging in free space. From the interference between $s$- and $d$-
partial waves in the differential scattering pattern we extract the
corresponding phase shifts. The method does not require knowledge of the
atomic density. This allows us to infer accurate values for the $s$- and
$d$-wave scattering amplitudes from the zero-energy limit up to the first
Ramsauer minimum using only the Van der Waals $C_{6}$ coefficient as
theoretical input. For the $^{87}$Rb triplet potential, the method reproduces
the scattering length with an accuracy of 6\%.

\end{abstract}

\maketitle

The scattering length $a$, the elastic scattering amplitude in the zero-energy
limit, is the central parameter in the theoretical description of quantum
gases \cite{Pitaevskii-03,Petrov-03,Fedichev-96}. It determines the kinetic
properties of these gases as well as the bosonic mean field. Its sign is
decisive for the collective stability of the Bose-Einstein condensed state.
Near scattering resonances, pairing behavior \cite{Petrov-03} and three-body
lifetime \cite{Fedichev-96} can also be expressed in terms of $a$. As a
consequence, the determination of the low-energy elastic scattering properties
is a key issue to be settled prior to further investigation of any new quantum gas.

Over the past decade the crucial importance of the scattering length has
stimulated important advances in collisional physics \cite{Weiner-99}. In all
cases except hydrogen \cite{Friend-80} the scattering length has to be
determined experimentally as accurate \textit{ab initio} calculations are not
possible. An estimate of the modulus $\left\vert a\right\vert $ can be
obtained relatively simply by measuring kinetic relaxation times
\cite{Monroe-93}. In some cases the sign of $a$ can be determined by such a
method, provided $p$-wave or $d$-wave scattering can be neglected or accounted
for theoretically \cite{Ferrari-02}. These methods have a limited accuracy
since they rely on the knowledge of the atomic density and kinetic properties.
Precision determinations are based on photo-association \cite{Heinzen-99},
vibrational-Raman \cite{Samuelis-01} and Feshbach-resonance spectroscopy
\cite{Chin-2000,Marte}, or a combination of those. They require refined
knowledge of the molecular structure in ground and excited electronic states
\cite{Weiner-99} .

In this Letter we present a stand-alone interference method for the accurate
determination of the full (\textit{i.e.} complex) $s$- and $d$-wave scattering
amplitudes in a quantum gas. Colliding two ultracold atomic clouds we observe
the scattering halo in the rest frame of the collisional center of mass by
absorption imaging. The clouds are accelerated up to energies at which the
scattering pattern shows the interference between the $s$- and $d$- partial
waves. After a computerized tomography transformation \cite{Born-Wolf} of the
images we obtain an angular distribution directly proportional to the
differential cross section. This allows us to measure the asymptotic phase
shifts $\eta_{l}(k)$ (with $k$ the relative momentum) of the $s$-wave $\left(
l=0\right)  $ and $d$-wave $\left(  l=2\right)  $ scattering channels. Using
these $\eta_{l}(k)$ as boundary conditions, we integrate the radial
Schr\"{o}dinger equation inwards over the $-C_{6}/r^{6}$ tail of the potential
and compute the accumulated phase \cite{AccumulatedPhase} of the wavefunction
at radius $20\,a_{0}$ (with $\,a_{0}$ the Bohr radius). All data of $\eta
_{l}(k)$ are used to obtain a single optimized accumulated phase from which we
can infer all the low-energy scattering properties, by integrating again the
same Schr\"{o}dinger equation outwards. Note that this procedure does not
require knowledge of the density of the colliding and scattered clouds, unlike
the stimulated raman detection approach of Ref.$\,$\cite{Legere-98}. We
demonstrate this method with $^{87}$Rb atoms interacting through the
ground-state triplet potential. We took data with both condensates and thermal
clouds. Here we report on the condensates, as they allow to observe the
largest range of scattering angles, $25{{}^{\circ}}<\theta<90{{}^{\circ}}$. Up
to $80\%$ of the atoms are scattered without destroying the interference
pattern. With our method, we obtain $a=+102(6)\,a_{0}$ for the scattering
length. The $d$-wave resonance \cite{Boesten-97} is found at the energy
$E_{res}=300(70)\;\mu$K. These results coincide within experimental error with
the precision determinations ($a=98.99(2)\,a_{0}$ \cite{Marte,Kempen-02} and
$E_{res}=270\;\mu$K \cite{Kempen-02}), obtained by combining the results of
several experiments as input for state-of-the-art theory.

In our experiments, we load about one billion $^{87}$Rb atoms in the (fully
stretched) $\left\vert F=2,m_{F}=2\right\rangle $ hyperfine level of the
electronic ground state from a magneto-optical trap (MOT) into a
Ioffe-Pritchard quadrupole trap ($21\times477$\thinspace Hz) with an offset
field of $B_{0}=+0.9\;$G. We pre-cool the sample to about $6\,\mu$K using
forced radio-frequency (RF) evaporation. The cloud is split in two by applying
a rotating magnetic field and ramping $B_{0}$ down to a negative value
$B_{0}^{-}$. This results in two Time-averaged Orbiting Potential (TOP) traps
loaded with atoms \cite{Tobias}. By RF-evaporative cooling we reach
Bose-Einstein condensation with about $10^{5}$ atoms in each cloud and a
condensate fraction of $\sim60\%.$

\begin{figure}[t]
\begin{center}
\includegraphics[height=8.5cm]{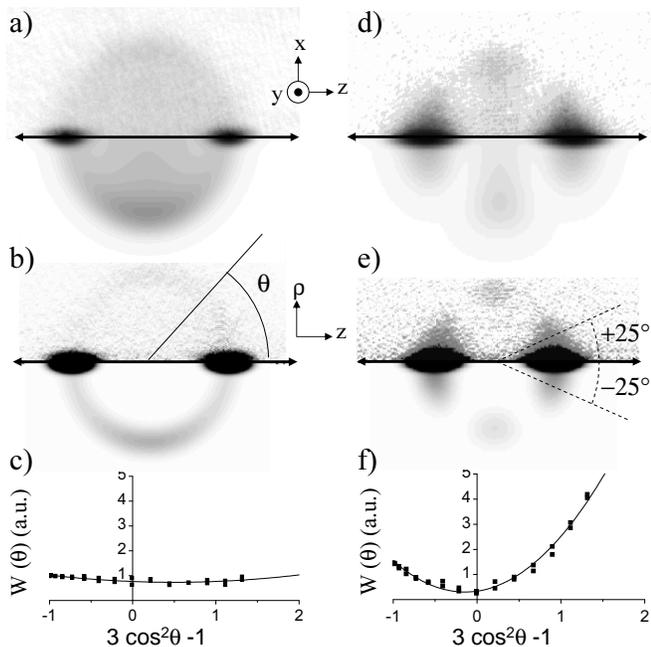}
\end{center}
\caption{a) Optical density of the scattering halo of two $^{87}$Rb
condensates for collision energy $E/k_{B}=138(4)\,\mu$K, measured 2.4 ms after
the collision (upper half: measured; lower half: calculated \cite{Calculation}
); b) radial density distribution obtained after tomography transformation of
image-a (upper half: measured; lower half: calculated \cite{Calculation}); c)
the dots show the angular scattering distribution $W\left(  \theta\right)  $
obtained after binning plot b, the line is the best parabolic fit. d,e,f) As
in plots a,b,c, but measured 0.5 ms after a collision at 1230(40)~$\mu$K. The
field of view of the images is $\sim1\times1$mm$^{2}$.}%
\label{buggle-1}%
\end{figure}

We then switch off the TOP fields and ramp $B_{0}$ back to positive values,
thus accelerating the clouds until they collide with opposite horizontal
momenta at the location of the trap center. The collision energies
$E=|2\mu_{B}B_{0}^{-}|=\hbar^{2}k^{2}/m$ (with $\mu_{B}$ the Bohr magneton and
$m$ the mass of $^{87}$Rb) range from $138\;\mu$K to $1.23\;$mK with an
overall uncertainty of $3\%$ (RMS). Approximately 0.5$\,$ms before the
collision we switch off the trap. A few ms later we observe the scattering
halo by absorption imaging. Fig.$\,$\ref{buggle-1}a (upper part) displays the
$s$-wave-dominated scattering halo (averaged over 20 pictures) of fully
entangled pairs (see \cite{Chikkatur-00}) obtained for a collision energy of
$E/k_{B}=138(4)$\ $\mu$K. In Fig.$\,$\ref{buggle-1}d (upper part), taken at
$E/k_{B}=1.23(4)\,$mK the halo is entirely different, showing a $d$%
-wave-dominated pattern. The lower halves of Fig.$\,$\ref{buggle-1}a and
Fig.$\,$\ref{buggle-1}d show the theoretical column densities $n_{2}\left(
x,z\right)  =\int n\left(  x,y,z\right)  dy$, where $n\left(  x,y,z\right)  $
is the calculated \cite{Calculation} density of the halo.

As the atoms are scattered by a central field, the scattering pattern must be
axially symmetric around the (horizontal) scattering axis ($z$-axis). As
pointed out by the Weizmann group \cite{Ozeri-02}, this allows a computerized
tomography transformation \cite{Born-Wolf} to reconstruct the radial density
distribution of the halo in cylindrical coordinates,
\begin{equation}
n\left(  \rho\mathbf{,}z\right)  =\frac{1}{4\pi}\int_{-\infty}^{\infty}
\tilde{n}_{2}\left(  \kappa_{x},z\right)  J_{0}\left(  \kappa_{x}\rho\right)
\left\vert \kappa_{x}\right\vert d\kappa_{x}.\label{TomgraphyTransform}%
\end{equation}
Here $\rho=\left(  x^{2}+y^{2}\right)  ^{1/2}$, $\tilde{n}_{2}\left(
\kappa_{x},z\right)  $ is the 1D Fourier transform along the $x$-direction of
the optical density with respect to $z$, and $J_{0}\left(  \varrho\right)  $
is the zero-order Bessel function. The transformed plots corresponding to the
images of Fig.$\,$\ref{buggle-1}a,d are shown as Fig.$\,$\ref{buggle-1}b,e respectively.

To obtain the angular scattering distribution $W(\theta)$ the tomography
pictures are binned in 40 discrete angular sectors. For gas clouds much
smaller than the diameter of the halo, $W\left(  \theta\right)  $ is directly
proportional to the differential cross section $\sigma\left(  \theta\right)
=2\pi\left\vert f\left(  \theta\right)  +f\left(  \pi-\theta\right)
\right\vert ^{2}$. Here, the Bose-symmetrized scattering amplitude is given by
a summation over the even partial waves, $f\left(  \theta\right)  +f\left(
\pi-\theta\right)  =(2/k)\sum_{l=even}(2l+1)e^{i\eta_{l}}P_{l}(\cos\theta
)\sin\eta_{l}$. Note that unlike in the \textit{total} elastic cross section
($\sigma=\int_{0}^{\pi/2}\sigma\left(  \theta\right)  \sin\theta d\theta
=(8\pi/k^{2})\sum_{l=even}(2l+1)\sin^{2}\eta_{l}$), the interference between
the partial waves is prominent in the \textit{differential} cross section.
Given the small collision energy in our experiments, only the $s$- and
$d$-wave scattering amplitudes contribute, $f_{s}\left(  \theta\right)
+f_{s}\left(  \pi-\theta\right)  =(2/k)e^{i\eta_{0}}\sin\eta_{0}$ and
$f_{d}\left(  \theta\right)  +f_{d}\left(  \pi-\theta\right)
=(2/k)(5/2)e^{i\eta_{2}}\left(  3\cos^{2}\theta-1\right)  \sin\eta_{2}$.
Therefore the differential cross section is given by
\begin{equation}
\sigma(\theta)=\frac{8\pi}{k^{2}}\sin^{2}\eta_{0}\left[  1+5\cos(\eta_{0}
-\eta_{2})u+\frac{25}{4}u^{2}\right]  ,\label{DifferentialCrossSection}%
\end{equation}
where $u\equiv\left(  \sin\eta_{2}/\sin\eta_{0}\right)  \left(  3\cos
^{2}\theta-1\right)  $.

To obtain the phase shifts, we plot the measured angular
distribution $W\left(  \theta\right)  $ as a function of $\left(
3\cos^{2}\theta-1\right) $ as suggested by
Eq.$\,$(\ref{DifferentialCrossSection}). The results for
Fig.$\,$\ref{buggle-1}a and Fig.$\,$\ref{buggle-1}d are shown as
the solid dots in Fig.$\,$\ref{buggle-1}c and
Fig.$\,$\ref{buggle-1}f, respectively. A parabolic fit to $W\left(
\theta\right)  $ directly yields a pair
$(\eta_{0}^{\exp}(k),\eta_{2}^{\exp}(k))$ of asymptotic phase
shifts (defined modulo $\pi$) corresponding to the two partial
waves involved \cite{PhaseShiftIndependent}. The absolute value of
$W\left( \theta\right) $ depends on quantities that are hard to
measure accurately (like the atom number) so we leave it out of
consideration. We rather emphasize that the measurement of the
phase shifts is a \textit{complete} determination of the (complex)
$s$- and $d$-wave scattering amplitudes at a given energy.

The radial wavefunctions corresponding to scattering at different
(low) collision energies and different (low) angular momenta
should all be in phase at small interatomic distances
\cite{AccumulatedPhase}. This so-called accumulated phase common
to all low-energy wavefunctions can be extracted from the full
data set $\{(\eta_{0}^{\exp}(k),\eta_{2}^{\exp}(k))\}$ mentioned
above. In practice, we use the experimental phase shifts
$\eta_{0}^{\exp}(k)$ and $\eta_{2}^{\exp }(k)$ as boundary
conditions to integrate inwards - for given $E$ and $l$ - the
Schr\"{o}dinger equation $\hbar^{2} d^{2}\chi(r)/dr^{2} +
p_{{}}^{2}(r) \chi (r) = 0$, and obtain the radial wavefunctions
$\chi(r)/r$ down to radius  $r_{in}=20\,a_{0}$. Here,
$p_{{}}^{2}(r)=m\left(E-V(r)\right)-\hbar^{2} l(l+1)/r^{2}$, where
$V(r)\simeq-C_{6}/r^{6}$ is the tail of the interaction potential.
At radius $20\,a_{0}$, the motion of the atoms is quasi-classical
and the accumulated phase can be written as $\Phi
(r)\simeq\arctan\left[ p(r)/(\hbar\,\partial\ln\chi/\partial
r)\right]  $.
\begin{figure}[t]
\begin{center}
\includegraphics[height=9cm]{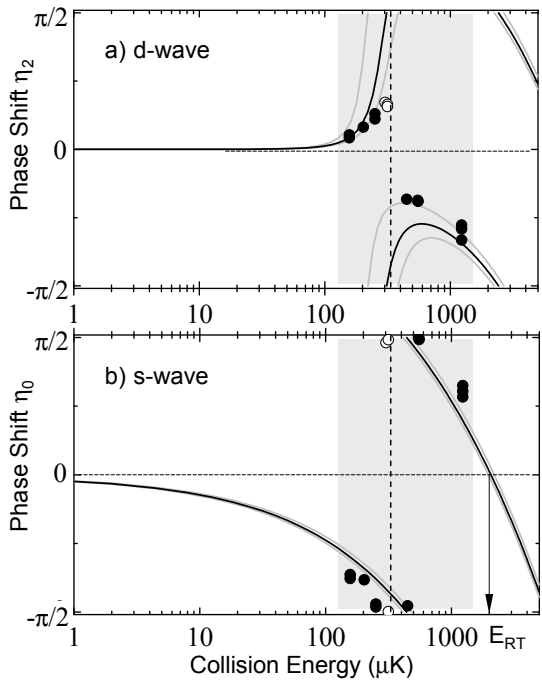}
\end{center}
\caption{a) $d$-wave and b) $s$-wave phase shifts versus collision
energy in $\mu$K. The circles are the results of the parabolic fit
of $W(\theta)$ for individual images. The full black lines is
calculated from the accumulated phase $\Phi_{\mathrm{opt}}$
optimized from all data points. The grey lines show the influence
of the uncertainty of $\pm\,\pi\times0.025$ on $\Phi
_{\mathrm{opt}}$. (The vertical dotted line indicates the
condition $\eta _{0}=\eta_{2}$). $s-d$ interference is only
observed in the gray areas. The first $s$-wave Ramsauer-Townsend
minimum is found at $E_{RT}=2.1(2)$ mK.} \label{PhaseShifts}
\end{figure}
The distance $20\,a_{0}$ is small enough \cite{InnerRadius} for
$\Phi(r_{in})$ to be highly insensitive to small variations in $E$
or $l$ \cite{AccumulatedPhase} and large enough that the
$-C_{6}/r^{6}$ part of the interaction potential is dominant over
the full range of integration. With a least-square method we
establish the best value $\Phi_{\mathrm{opt}}
(r_{in})=1.34\pm\pi\times0.025$ for the accumulated phase at
$20\,a_{0}$ \cite{LeastSquare}. Here the error bar reflects the
experimental accuracy and not the systematic error related to the
choice of $C_{6}$, the latter being of less relevance as discussed
below. Interestingly, the $d$-wave scattering resonance
\cite{Boesten-97} results in a sudden variation of
$\eta_{2}^{\exp}$ with the collision energy in the vicinity of
that resonance (see Fig.$\,$ \ref{PhaseShifts}a). This imposes a
stringent condition on the optimization of $\Phi_{\mathrm{opt}}$
and constrains its uncertainty.

Once $\Phi_{\mathrm{opt}}$ has been established, one can use it as a boundary
condition to integrate the Schr\"{o}dinger equation outwards and compute
$\eta_{l}(k)$ for any desired (low) value of $k$ and $l$. Fig.$\,$%
\ref{PhaseShifts} shows the resulting phase shifts for collision
energies up to $5$ mK \cite{g-wave}.  The first Ramsauer-Townsend
minimum \cite{RamsauerMinimum} in the $s$-wave cross section is
found at collision energy $E_{RT}/k_{B}=2.1(2)$ mK. The solid dots
represent the $\eta_{l}^{\exp}(k_{i})$ obtained from the parabolic
fit of $W\left(  \theta\right)  $ from individual images. The
three open circles correspond to measurements for which the sign
of the phase shifts could not be established \cite{Signs}.
Refinements to the present data analysis may include the
occurrence of multiple scattering as well as the influence of the
spatial extension of the colliding clouds taking into account the
non condensed fraction.

\begin{figure}[t]
\begin{center}
\includegraphics[height=5cm]{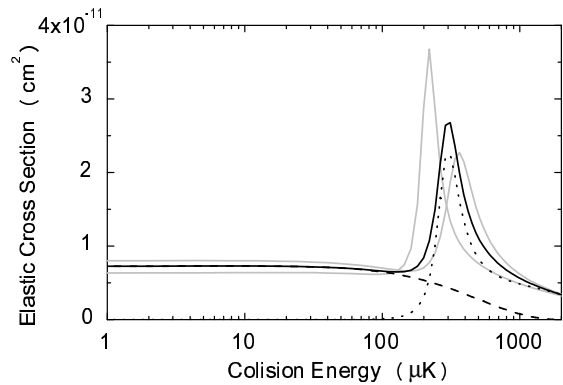}
\end{center}
\caption{s-wave (dashed line), d-wave (dotted line) and total
(full black line) elastic cross sections (in cm$^{2}$) versus
collision energy (in $\mu $K), computed from the optimized
accumulated phase $\Phi_{\mathrm{opt}}$ as determined in this
work. The gray lines are the total elastic cross sections,
obtained from $\Phi_{\mathrm{opt}} \pm \pi \times0.025 $.}
\label{Fig: CrossSection}
\end{figure}

Knowing the phase shifts, we can infer all the low-energy scattering
properties. Our results for the elastic scattering cross section are shown in
Fig.$\,$\ref{Fig: CrossSection}. The (asymmetric) $d$-wave resonance emerges
pronouncedly at $300(70)\ \mu$K with an approximate width of $150\;\mu$K
(FWHM). Most importantly, the scattering length follows from the
$k\rightarrow0$ limiting behavior, $\eta_{0}(k\rightarrow0)=-ka$. We find
$a=+102(6)\,a_{0}$, whereas the state-of-the-art value is $a=98.99(2)\,a_{0}$
\cite{Kempen-02}.

Comparison with the precision determinations
\cite{Marte,Kempen-02} shows that our method readily yields fairly
accurate results, relying only on input of the $C_{6}$
coefficient. We used the value $C_{6}=4.698(4)\times10^{3}\;$a.u.
\cite{Kempen-02}. In the present case, one does not need to know
$C_{6}$ to this accuracy. Increasing $C_{6}$ by $10\%$ results in
a 1 $\%$-change of the scattering length. Clearly, the systematic
error in $\Phi_{\mathrm{opt}}$ accumulated by integrating the
Schr\"{o}dinger equation inward with a wrong $C_{6}$ largely
cancels when integrating back outward. However, in the case of a
$s$-wave resonance other atomic species may reveal a stronger
influence of $C_{6}$ on the calculated scattering length. Simple
numerical simulations show that the value of $C_{6}$ becomes
critical only when the (virtual) least-bound state in the
interaction potential has an extremely small (virtual) binding
energy (less than $10^{-2}$ level spacing). Hence our method
should remain accurate in almost any case.

This method can therefore be applied to other bosonic or fermionic
atomic species, provided the gases can be cooled and accelerated
in such a way that the lowest-order partial-wave interference can
be observed with good energy resolution. We speculate that the
accuracy of the method can be strongly improved by turning to
smaller optical-density clouds and fluorescence detection. It will
enable higher collision energies and observation of higher-order
partial-wave interference. The use of more dilute clouds and
longer expansion times will also eliminate multiple-scattering
effects and finite-size convolution broadening of the interference
pattern. Finally it will enable precision measurements of the
scattered fraction, which in the case of $^{87}$Rb will allow us
to pinpoint the location of the $d$-wave resonance to an accuracy
of $10\,\mu$K or better. In combination with state-of-the-art
theory such improvements are likely to turn our approach into a
true precision method.

Similar experiments were reported during the final stage of completion of this
Letter \cite{Kiwis}.

The authors acknowledge valuable discussions with S. Kokkelmans,
D. Petrov, G. Shlyapnikov, S. Gensemer and B. Verhaar. This work
is part of the research programme of the `Stichting voor
Fundamenteel Onderzoek der Materie (FOM), supported by the
`Nederlandse organisatie voor Wetenschappelijk Onderzoek (NWO)'.
JL acknowledges support from a Marie Curie Intra-European
Fellowship (MEIF-CT-2003-501578).


\begin{thebibliography}{99}
\bibitem {InnerRadius}We checked that, within the range $15\,a_{0}%
<r<25\,a_{0}$, the exact choice of the inner radius of the integration
interval is of no influence for the results presented in this Letter.
\bibitem{LeastSquare} The determination of $\Phi_{\mathrm{opt}}$ is done by minimizing $\left[
{\textstyle\sum_{i=1}^{N}}{\textstyle\sum_{l=0,2}} \left(
\eta_{l}^{\exp}(k_{i})-\eta_{l}^{\mathrm{opt}}(k_{i})\right)
^{2}\right]  ^{1/2}$, where N is the number of data points, and
the $\eta_{l}^{\mathrm{opt}}(k_{i})$ are the phase shifts computed
from $\Phi_{\mathrm{opt}}$.
\bibitem {g-wave}The $g$-wave $(l=4)$ phase shift can also be computed from
the same $\Phi_{\mathrm{opt}}$, but the error made by assuming a constant
accumulated phase increases like $l(l+1)$, and the resulting $g$-wave would be
accordingly less accurate.
\bibitem {RamsauerMinimum}Ramsauer-Townsend minima are observed around
collision energies where a phase shift crosses zero (see
Fig.\thinspace \ref{PhaseShifts}).
\bibitem {Signs}Since Eq.$\,$(\ref{DifferentialCrossSection}) is unchanged
under reversal of the sign of both $\eta_{0}^{\exp}(k)$ and
$\eta_{2}^{\exp }(k)$, the result of the parabolic fit is not
unique. We eliminate this ambiguity by comparing the difference in
accumulated phases obtained for $\eta_{0}^{\exp}(k)$ and
$\eta_{2}^{\exp}(k)$ with that obtained for $-\eta_{0}^{\exp}(k)$
and $-\eta_{2}^{\exp}(k)$. We find that in almost all cases, this
difference - which should be vanishingly small - is much larger
for one choice of signs than for the other. We conclude that the
correct sign for a pair $(\eta _{0}^{\exp}(k),\eta_{2}^{\exp}(k))$
is the one for which the difference between the accumulated phases
is the smallest. In all but three cases (at the same collision
energy), this criterium is conclusive. These three measurements
(the open circles in Fig.$\,$\ref{PhaseShifts}) are left out of
the procedure used to compute $\Phi_{\mathrm{opt}}$. In hindsight
they turn out to correspond to the marginal case mentioned in
\cite{PhaseShiftIndependent}.
\end{thebibliography}
\end{document}